\documentclass[a4paper,11pt]{article}
\usepackage{amsmath}        
\usepackage{amssymb}        
\usepackage{cite}
\usepackage{cancel}
\usepackage{fancybox}
\usepackage{color}
\usepackage{ulem}
\usepackage{pifont}
\usepackage{fancyhdr}
\usepackage{longtable}

\def\USp{U\hspace{-0.125em}Sp}

\newsavebox{\BCmatrixA}
\savebox{\BCmatrixA}{$\left(\begin{array}{cc}\eta_2&\eta_3\\ \eta_0&\eta_1\end{array}\right)$}
\newsavebox{\BCmatrixB}
\savebox{\BCmatrixB}{$\left(\begin{array}{c}\eta_2\\ \eta_0\end{array}\right)$}
\newsavebox{\BCmatrixC}
\savebox{\BCmatrixC}{$\left(\begin{array}{cc}\eta_{2\pm}\overline{\gamma}P_{2{\bf 32}}&\eta_{3\pm}\overline{\gamma}P_{3{\bf 32}}\\ \eta_{0\pm}\overline{\gamma}P_{0{\bf 32}}&\eta_{1\pm}\overline{\gamma}P_{1{\bf 32}}\end{array}\right)$}
\newsavebox{\BCmatrixD}
\savebox{\BCmatrixD}{$\left(\begin{array}{cc}\eta_2P_{2{\bf 32}}TP_{2{\bf 32}}^{-1}&\eta_3P_{3{\bf 32}}TP_{3{\bf 32}}^{-1}\\ \eta_0P_{0{\bf 32}}TP_{0{\bf 32}}^{-1}&\eta_1P_{1{\bf 32}}P_{1{\bf 32}}^{-1}\end{array}\right)$}

\usepackage{diagbox}
\usepackage{lscape}
\usepackage{colortbl}
\usepackage{multicol}
\makeatletter
 
 \@addtoreset{equation}{section}
\makeatother

\makeatletter
 
 \@addtoreset{equation}{section}
\makeatother
\begin{document}
\thispagestyle{fancy}

\title{$\USp(32)$ Special Grand Unification}

\author{Naoki Yamatsu
\footnote{Electronic address: yamatsu.naoki@phys.kyushu-u.ac.jp}
\\
{\it\small Department of Physics, Kyushu University, Fukuoka 819-0395, Japan}
}
\date{\today}

\maketitle

\thispagestyle{fancy}

\begin{abstract}
 We discuss a grand unified theory (GUT) based on a $\USp(32)$ GUT gauge
 group broken to its subgroups including a special subgroup.
 A GUT based on an $SO(32)$ GUT gauge group has
 been discussed on six-dimensional (6D) orbifold space
 $M^4\times T^2/\mathbb{Z}_2$.
 It is inspired by the $SO(32)$ string theory behind the $SU(16)$ GUT
 whose $SU(16)$ is broken to a special subgroup $SO(10)$. 
 Alternative direction is to embed an $SU(16)$ gauge group into a
 $\USp(32)$ GUT gauge group, which is inspired by a 
 non-supersymmetric symplectic-type $\USp(32)$ string theory.
 In a $\USp(32)$ GUT, one generation of the SM fermions is
 embedded into a 6D bulk Weyl fermion in a $\USp(32)$ defining
 representation.
 For a three generation model, all the 6D and 4D gauge
 anomalies in the bulk and on the fixed points are canceled out without
 exotic chiral fermions at low energies.
 The SM Higgs scalar is embedded into a 6D bulk scalar field in
 a $\USp(32)$ adjoint representation.
\end{abstract}

\section{Introduction}
\label{Sec:Introduction}

The Standard Model (SM) in particle physics explains almost all
experiments and observations at low energies, but the SM is usually
regarded as a low-energy effective theory. There are many attempts
to construct unified theories beyond the SM as below.

Grand unification \cite{Georgi:1974sy} is one of the most attractive
ideas to construct unified theories beyond the SM.
As is well-known in e.g., Refs.~\cite{Slansky:1981yr,Yamatsu:2015gut},
in four-dimensional (4D) space-time framework, the candidates for grand
unified theory (GUT) gauge groups  are only
$SU(n) (n\geq 5)$\cite{Georgi:1974sy,Inoue:1977qd},
$SO(4n+2) (n\geq 2)$\cite{Fritzsch:1974nn,Ida:1980ea,Fujimoto:1981bv},
and $E_6$\cite{Gursey:1975ki}
because of ranks of groups and types of representations.
In higher dimensional space-time frameworks \cite{Yamatsu:2015gut},
additional Lie groups such as $SO(11)$
\cite{Hosotani:2015hoa,Yamatsu:2015rge,Furui:2016owe,Hosotani:2017ghg,Hosotani:2017edv,Englert:2019xhz,Englert:2020eep}
$SO(12)$
\cite{Nomura:2008sx,Chiang:2011sj},
and $E_{7,8}$
are also candidates for GUT gauge groups.
A lot of GUT models have been already proposed based on GUT gauge groups
which are broken only to {\it regular subgroups};
e.g., 
\begin{align}
 E_8\underset{(R)}{\longrightarrow}
 E_7\underset{(R)}{\longrightarrow}
 E_6\underset{(R)}{\longrightarrow}
 SO(10)
 \underset{(R)}{\longrightarrow}
 \left\{
 \begin{array}{c}
  SU(5)\\
  \mbox{or}\\
  G_{\rm PS}\\
 \end{array}
 \right\}
 \underset{(R)}{\longrightarrow} G_{\rm SM},
\end{align}
where the subscript of arrows $(R)$ stands for a regular subgroup
breaking, 
$G_{\rm SM}:=SU(3)_C\times SU(2)_L\times U(1)_Y$
stands for the SM gauge group,
$G_{\rm PS}:=SU(2)_L\times SU(2)_R\times SU(4)_C$
stands for the Pati-Salam group \cite{Pati:1974yy},
and we omitted several $U(1)$ subgroups.
A few GUT models \cite{Yamatsu:2017sgu,Yamatsu:2017ssg,Yamatsu:2018fsg}
have been already proposed based on GUT gauge groups
which are broken not only to regular subgroups
but also to {\it special subgroups}; e.g.,
\begin{align}
 SO(32)
 \underset{(R)}{\longrightarrow} SU(16)
 \underset{(S)}{\longrightarrow} SO(10)
 \underset{(R)}{\longrightarrow}
 \left\{
 \begin{array}{c}
  SU(5)\\
  \mbox{or}\\
  G_{\rm PS}\\
 \end{array}
 \right\}
 \underset{(R)}{\longrightarrow} G_{\rm SM},
\label{Eq:Symmetry-breaking_SO32-GSM} 
\end{align}
where the subscripts of arrows $(R)$ and $(S)$ stand for regular and
special subgroup breakings, respectively,
and we omitted several $U(1)$ subgroups for regular subgroups.
We note that a subgroup $H$ of a group $G$ is called a regular subgroup if
all the Cartan subgroups of $H$ are also the Cartan subgroups of $G$; 
otherwise, the subgroup $H$ is called a special subgroup
\cite{Dynkin:1957um,Dynkin:1957ek}.
\footnote{
For Lie groups and their regular and special subgroups, see e.g.,
Refs.~\cite{Dynkin:1957ek,Dynkin:1957um,Cahn:1985wk,Slansky:1981yr,Yamatsu:2015gut}.
Branching rules of Lie groups and their subgroups such as
$SO(32)\supset SU(16)$ and $SU(16)\supset SO(10)$ are explicitly
written in Ref.~\cite{Yamatsu:2015gut}. Other branching rules used
in the letter are calculated by a Mathematica program Susyno
\cite{Fonseca:2011sy} and each projection matrix of Lie groups and their
subgroups obtained by the wight diagram method discussed 
in e.g., Refs.~\cite{Cahn:1985wk,Yamatsu:2015gut}.}

In Ref.~\cite{Yamatsu:2017sgu},
the author proposed a GUT model based on an $SU(16)$ GUT gauge group
broken to a non-maximal subgroup $G_{\rm SM}$ shown in
Eq.~(\ref{Eq:Symmetry-breaking_SO32-GSM}), where this type of GUTs 
are referred as special GUTs below. The results are summarized as
follows. In a 4D $SU(16)$ special GUT, one generation of quarks and
leptons is embedded into a 4D $SU(16)$ ${\bf 16}$ Weyl fermion because
one generations of the SM Weyl fermion fields are correctly unified
into a Weyl fermion field in an $SO(10)$ spinor representation
${\bf 16}$.
Only three 4D $SU(16)$ ${\bf 16}$ Weyl fermions suffer from a 4D
$SU(16)$ gauge anomaly \cite{Banks1976}.
It is possible to cancel out the anomaly by introducing three 4D
$SU(16)$ ${\bf \overline{16}}$ Weyl fermions,
where a representation ${\bf \overline{R}}$ is the complex conjugate of
${\bf R}$, but the matter content is
vectorlike, and it is far from the SM whose matter content is chiral.
So, to satisfy the 4D anomaly cancellation in a chiral gauge theory, we
need to introduce a 4D Weyl fermion that belongs to
a representation ${\bf R}(\not={\bf \overline{16}})$.
One of the candidates is a 4D Weyl fermion in the $SU(16)$
anti-symmetric tensor representation ${\bf 120}$,
where the value of the 4D anomaly coefficient of the $SU(16)$
anti-symmetric tensor representation ${\bf 120}$ is twelve times greater
than one of the $SU(16)$ defining representation ${\bf 16}$.
That is, the anomaly coefficient of the 4D $SU(16)$
$(12\times {\bf 16}\oplus{\bf \overline{120}})$ is zero, so
4D $SU(16)$ $(12\times {\bf 16}\oplus{\bf \overline{120}})$ Weyl
fermions satisfy the 4D $SU(16)$ anomaly cancellation condition, while
the 4D $SU(16)$ $(12\times {\bf 16}\oplus{\bf \overline{120}})$ Weyl
fermions are identified as twelve generations of quarks and leptons
at a vacuum. We note that since a complex representation
${\bf \overline{120}}$ of $SU(16)$ is identified with a real
representation ${\bf 120}$ of $SO(10)$,
a 4D $SU(16)$ ${\bf \overline{120}}$ Weyl fermion is not a chiral
fermion at a vacuum whose $SU(16)$ is broken to $SO(10)$.
Regardless of a choice of a Weyl fermion in a representation ${\bf R}$
of $SU(16)$, a 4D $SU(16)$ gauge anomaly cancellation condition
restricts the minimal number of generations.
Unfortunately, the minimal number is 12 in a 4D framework.
In a 6D $SU(16)$ special GUT on 6D orbifold space 
$M^4\times T^2/\mathbb{Z}_2$, which is 4D
Minkowski spacetime and two extra dimensional compactified space $T^2$
with a $\mathbb{Z}_2$ orbifold structure, 
one generation of quarks and leptons is identified as
a set of the zero modes of a 6D $SU(16)$ ${\bf 16}$ Weyl fermion.
Three generations of quarks and leptons are allowed without 4D exotic
chiral fermions, 
which is consistent with the $SU(16)$ gauge anomaly cancellation in the
bulk and  on the fixed points.

Further, an $SO(32)$ special GUT
was proposed in Ref.~\cite{Yamatsu:2017ssg} 
whose gauge group $SO(32)$ is broken to
a non-maximal subgroup $G_{\rm SM}$ shown in
Eq.~(\ref{Eq:Symmetry-breaking_SO32-GSM}).
Since an $SO(32)$ group has only real representations
\cite{Slansky:1981yr,Yamatsu:2015gut}, any 4D $SO(32)$ gauge theory is a
vectorlike theory. 
To realize the SM, i.e., a 4D chiral gauge theory, 
an orbifold space construction 
\cite{Kawamura:1999nj,Kawamura:2000ev,Asaka:2001eh,Kawamura:2009sa,Kawamura:2009gr}
is used.
The results of the $SO(32)$ special GUT on 6D orbifold space
$M^4\times T^2/\mathbb{Z}_2$ are almost the same as the above $SU(16)$
special GUT.

The $SO(32)$ special GUT was inspired by superstring theories
\cite{Polchinski1998a,Polchinski1998b}
that have been considered as a candidate of a unified theory to describe
all the interaction including gravity. 
There are a lot of attempts to construct the SM as an effective theory
derived from supersymmetric $E_8\times E_8$ and $SO(32)$ 
string theories \cite{Green1984,Aldazabal:1994zk,Abe:2015mua,Abe:2015xua,Abe:2016eyh,Nibbelink:2016wms,Otsuka:2018oyf,Fraiman:2018ebo,Otsuka:2018rki}. 
From the low-energy experiments, even if more fundamental theories
beyond the SM have supersymmetry, the supersymmetry must be broken.
So, it may be worth considering not only supersymmetric but also
non-supersymmetric string theories.

Non-supersymmetric string theories has been investigated in 
e.g., Refs.~\cite{Sen:1999md,Yoneya:1999qe,Sugimoto:1999tx,Schwarz:2001sf,Moriyama:2001ge,Angelantonj:2007ts,Blaszczyk:2014qoa,Angelantonj:2015nfa,Abel:2015oxa,Blaszczyk:2015zta}.
It was shown in Ref.~\cite{Sugimoto:1999tx}
that a non-supersymmetric
symplectic-type $\USp(32)$ string theory satisfies the gravitational and
gauge anomaly cancellation conditions, which is further investigated in
e.g., Refs.~\cite{Schwarz:2001sf,Moriyama:2001ge,Angelantonj:2007ts}.
Several attempts to construct the SM derived from non-supersymmetric
$SO(16)\times SO(16)$ string theories exist
\cite{Blaszczyk:2014qoa,Angelantonj:2015nfa,Abel:2015oxa,Blaszczyk:2015zta},
but those from a non-supersymmetric $\USp(32)$ string theory 
do not exist.
In this letter, we propose a $\USp(32)$ model, but we will not discuss
how to realize the model derived from string theories.

From another viewpoint of unified theories beyond the SM,
gauge-Higgs unification (GHU)
\cite{Hosotani:1983xw,Hosotani:1988bm,Davies:1987ei,Davies:1988wt,Hatanaka:1998yp,Hatanaka:1999sx}
is also an interesting idea, where the Higgs boson is identified as 
a zero mode of extra dimensional components of higher
dimensional gauge fields.
There are many GHU models discussed in e.g., 
Refs.~\cite{Kawamura:1999nj,Kawamura:2000ir,Kawamura:2000ev,Burdman:2002se,Lim:2007jv,Kojima:2011ad,Kojima:2016fvv,Kubo:2001zc,Csaki:2002ur,Scrucca:2003ra,Khojali:2017azj}.
Realistic GHU models are based on $SU(3)_C \times SO(5) \times U(1)$
gauge theories in the Randall-Sundrum (RS) warped spacetime
\cite{Agashe:2004rs,Medina:2007hz,Hosotani:2008tx,Funatsu:2013ni,Funatsu:2014fda,Funatsu:2016uvi,Yoon:2018vsc,Yoon:2018xud,Funatsu:2019xwr,Funatsu:2019fry,Funatsu:2020znj,Funatsu:2020haj}.
It gives nearly the same phenomenology at low energies as the SM.
One of the $SU(3)_C \times SO(5) \times U(1)$ GHU models is embedded in 
an $SO(11)$ GHU GUT model
\cite{Hosotani:2015hoa,Yamatsu:2015rge,Furui:2016owe,Hosotani:2017edv,Hosotani:2017ghg,Englert:2019xhz,Englert:2020eep},
where the SM gauge group and matter content are incorporated into
GUT in five or six dimensional RS warped spacetime.
Other GHU GUT models based on, e.g., 
$SU(6)$ \cite{Burdman:2002se,Lim:2007jv,Maru:2019lit,Maru:2019bjr},
$SO(12)$ \cite{Nomura:2008sx,Chiang:2011sj},
$E_6$ \cite{Chiang:2010hy,Kojima:2017qbt}
are also discussed.
As is known in e.g., Refs.~\cite{Slansky:1981yr,Yamatsu:2015gut}, 
an $SO(11)$ spinor representation ${\bf 32}$ is a pseudo-real
representation, so ${\bf 32}$ of $SO(11)$ cannot be identified as
${\bf 32}$ of $SO(32)$, while ${\bf 32}$ of $SO(11)$ can be identified
as ${\bf 32}$ of $\USp(32)$. We note that $SO(11)$ is not a maximal
subgroup of $\USp(32)$, and $SO(12)(\supset SO(11))$ is a maximal
subgroup of $\USp(32)$.
In this letter, we propose a $\USp(32)$ model, but it is not a
gauge-Higgs GUT model.

In the letter, we propose a $\USp(32)$ special GUT on 6D orbifold
spacetime $M^4\times T^2/\mathbb{Z}_2$ whose $\USp(32)$ gauge group is
broken to $G_{\rm SM}$ by orbifold symmetry breaking and the Higgs
mechanism. 
We mainly focus on how to realize the SM matter content.
It is an important task for constructing unified theories beyond the SM
based on symplectic groups because almost people seems to believe that
symplectic groups cannot be applied for GUT model buildings. 
The model building of a $\USp(32)$ special GUT is almost parallel
to an $SO(32)$ special GUT discussed in Ref.~\cite{Yamatsu:2017ssg}.
As we see below, the results of the $\USp(32)$ special GUT
are almost the same as the ones of
the $SO(32)$ one.

This letter is organized as follows. In Sec.~\ref{Sec:Special-GUT}, 
we construct a 6D $\USp(32)$ special GUT on $M^4\times T^2/\mathbb{Z}_2$.
Section~\ref{Sec:Summary-discussion} is devoted to a summary and
discussion.

\section{$\USp(32)$ special GUT}
\label{Sec:Special-GUT}

Before we introduce a $\USp(32)$ special GUT, we quickly check the
$\USp(32)$ group and its subgroups. 
One of the maximal subgroups of a $\USp(32)$ group is a
regular subgroup $SU(16)\times U(1)$.
The branching rules of 
$\USp(32)\supset SU(16)\times U(1)$ for 
$\USp(32)$ defining, rank-2 anti-symmetric tensor, and
rank-2 symmetric tensor (adjoint) representations ${\bf 32}$,
${\bf 495}$, and ${\bf 528}$
are given by
\begin{align}
{\bf 32}&=
({\bf 16})(1)
\oplus({\bf \overline{16}})(-1),
\label{branching-rule-32-USp32-SU16}\\
{\bf 495}&=
({\bf 255})(0)
\oplus({\bf 120})(2)
\oplus({\bf \overline{120}})(-2),
\label{branching-rule-496-USp32-SU16}\\
{\bf 528}&=
({\bf 255})(0)
\oplus({\bf 1})(0)
\oplus({\bf 136})(2)
\oplus({\bf \overline{136}})(-2),
\label{branching-rule-527-USp32-SU16}
\end{align}
where the branching rules of 
$SU(16)\supset SO(10)$ for $SU(16)$ defining, rank-2 anti-symmetric
tensor, rank-2 symmetric tensor, and adjoint representations
${\bf 16}$ $ ({\bf \overline{16}})$,
${\bf 120}$ $({\bf \overline{120}})$,
${\bf 126}$ $({\bf \overline{126}})$, and
${\bf 255}$
are given in Ref.~\cite{Yamatsu:2015gut} as
\begin{align}
&{\bf 16}=({\bf 16}),\ \
({\bf \overline{16}})=({\bf \overline{16}}),
\label{branching-rule-16-SU16-SO10}\\
&{\bf 120}=({\bf 120}),\ \
{\bf \overline{120}}=({\bf 120}),
\label{branching-rule-120-SU16-SO10}\\
&{\bf 136}=({\bf \overline{126}})\oplus({\bf 10}),\ \
{\bf \overline{136}}=({\bf 126})\oplus({\bf 10}),
\label{branching-rule-136-SU16-SO10}\\
&{\bf 255}=
({\bf 210})
\oplus({\bf 45}).
\label{branching-rule-255-SU16-SO10}
\end{align}
From Eqs.~(\ref{branching-rule-32-USp32-SU16}) and
(\ref{branching-rule-16-SU16-SO10}),
a $\USp(32)$ defining representation is decomposed into
$SO(10)$ spinor representations ${\bf 16}$ and ${\bf \overline{16}}$.
From Eqs.~(\ref{branching-rule-527-USp32-SU16}),
(\ref{branching-rule-136-SU16-SO10}), and 
(\ref{branching-rule-255-SU16-SO10}),
a $\USp(32)$ adjoint representation ${\bf 528}$
is decomposed into $SO(10)$ bi-spinor ${\bf 210}$,
adjoint ${\bf 45}$,
rank-2 symmetric tensor ${\bf 126}$ and ${\bf \overline{126}}$, and 
two vector representations ${\bf 10}$.
In many $SO(10)$ GUT models
\cite{Fritzsch:1974nn,Aulakh:1982sw,Babu:1992ia,Aulakh:2003kg,Bajc:2005zf,Fukuyama2005,Bertolini:2009qj,Altarelli:2013aqa,Fukuyama:2012rw,Mambrini:2015vna,Ellis:2018khn,Ferrari:2018rey,Chakrabortty:2019fov,Chakraborty:2019uxk},
not only vector, spinor and adjoint  representations
${\bf 10}$, ${\bf 16}$ (${\bf \overline{16}}$) and ${\bf 45}$ but also
rank-2 symmetric tensor and bi-spinor representations
${\bf 126}$ (${\bf \overline{126}}$) and ${\bf 210}$
are introduced.
For example,
an $SO(10)$ ${\bf 126}$ (${\bf \overline{126}}$) scalar
field is introduced to generate neutrino masses via a 4D
renormalizable cubic term \cite{Babu:1992ia,Aulakh:2003kg},
to reproduce realistic Yukawa coupling constants of quarks and leptons
\cite{Aulakh:2003kg,Bajc:2005zf},
to realize gauge coupling unification
\cite{Bajc:2005zf,Bertolini:2009qj},
and to introduce a dark matter candidate
\cite{Mambrini:2015vna,Ferrari:2018rey};
an $SO(10)$ ${\bf 210}$ scalar field is also introduced to break $SO(10)$
to $G_{\rm PS}$ and to realize gauge coupling unification
\cite{Bajc:2005zf,Bertolini:2009qj},
to introduce a dark matter candidate
\cite{Mambrini:2015vna,Ferrari:2018rey},
and to prevent rapid proton decay
\cite{Chakrabortty:2019fov,Chakraborty:2019uxk}.

We propose a $\USp(32)$ special GUT in a 6-dimensional (6D) hybrid
warped space $M_4\times T^2/\mathbb{Z}_2$. The metric is given by 
\cite{Randall:1999ee,Hosotani:2017ghg}
\begin{align}
ds^2=e^{-2\sigma(y)}(\eta_{\mu\nu}dx^{\mu}dx^{\nu}+dv^2)+dy^2,
\label{Eq:6D-RS-metric}
\end{align}
where $\eta_{\mu\nu}=\mbox{diag}(-1,+1,+1,+1)$, 
$\sigma(y)=\sigma(-y)=\sigma(y+2\pi R_5)$,
$\sigma(y)=k|y|$ for $|y|\leq \pi R_5$,
and $R_5$ and $R_6$ represent 5th and 6th spatial extra dimensional 
radiuses, respectively.
We identify spacetime points
$(x^\mu,y,v), (x^\mu,y+2\pi R_5,v), (x^\mu,y,v+2\pi R_6)$, and
$(x^\mu,-y,-v)$. 
There are four fixed points in the extra-dimensional space under
$\mathbb{Z}_2$ parity:
$(y_0,v_0)=(0,0)$, $(y_1,v_1)=(\pi R_5,0)$, 
$(y_2,v_2)=(0,\pi R_6)$, and $(y_3,v_3)=(\pi R_5,\pi R_6)$.
Parity $P_j$ $(j=1,2,3,4)$ around each fixed point is defined by
$(x_\mu,y_j+y,v_j+v)\ \to\ (x_\mu,y_j-y,v_j-v)$,
where $P_3=P_1P_0P_2=P_2P_0P_1$.
5th and 6th dimensional translations
$U_5: (x_\mu,y,v)\to(x_\mu,y+2\pi R_5,v)$ 
and $U_6: (x_\mu,y,v)\to(x_\mu,y,v+2\pi R_6)$  
satisfy $U_5=P_1P_0$ and $U_6=P_2P_0$, respectively.
The metric given in Eq.~(\ref{Eq:6D-RS-metric})
becomes a solution of the Einstein
equations with the brane tension at $y=0$ and $\pi R_5$ and a negative
cosmological constant $\Lambda=-10k^2$. There are 4D branes at
$y=0$ and $\pi R_5$ referred to as the ultra-violet (UV) and infra-red
(IR) branes, respectively.

We introduce three different dimensional fields in the $\USp(32)$
special GUT.
First, we introduce a 6D bulk gauge field $A_M$, which contains
the SM gauge fields, where
the gauge field belongs to the adjoint representation of $\USp(32)$
${\bf 528}$;
a 6D bulk scalar field $\Phi_{\bf 528}$, which contains
the SM Higgs field, where the subscript of fields, e.g., {\bf 528}
stands for the representation of $\USp(32)$;
three sets of 6D bulk positive and negative Weyl fermions
$\Psi_{{\bf 32}+}^{(a)}$ and $\Psi_{{\bf 32}-}^{(a)}$ $(a=1,2,3)$,
which contains the SM fermion fields.
Second, to realize symmetry breaking at low energies, we introduce
three 5D $\USp(32)$ ${\bf 86800}$, ${\bf 495}$, and ${\bf 32}$ brane
scalar fields $\Phi_{\bf 86800}$, $\Phi_{\bf 495}$, and $\Phi_{\bf 32}$
on the UV brane.
Third, to realize 4D anomaly cancellation on all the fixed points,
we introduce
$\psi_{\overline{\bf 120}}$,
$\psi_{{\bf 16}}^{(b)}$, $\psi_{\overline{\bf 16}}^{(b)}$
$(b=1,2,\cdots,12)$
at $(y,v)=(y_0,v_0)$, 
where the subscript of fields, e.g., {\bf 120}
stands for the representation of $SU(16)$.
The matter content of the $\USp(32)$ special GUT is summarized in
Table~\ref{tab:USp32-SU16-SO10-matter-content-6D}. 

\begin{table}[t]
\begin{center}
\begin{tabular}{ccccc}\hline
\rowcolor[gray]{0.8}
6D Bulk field&
$A_M$&$\Phi_{{\bf 528}}$&$\Psi_{{\bf 32}+}^{(a)}$&$\Psi_{{\bf 32}-}^{(a)}$
\\\hline
$\USp(32)$ &${\bf 528}$&${\bf 528}$&${\bf 32}$&${\bf 32}$\\
$SO(5,1)$&${\bf 6}$&${\bf 1}$&${\bf 4}_+$&${\bf 4}_-$\\
Orbifold BC&
&$\left(
\begin{array}{cc}
+&+\\
-&-\\
\end{array}
 \right)$
&$\left(
\begin{array}{cc}
-&-\\
-&-\\
\end{array}
\right)$
&$\left(
\begin{array}{cc}
+&-\\
+&-\\
\end{array}
\right)$
\\
\hline
\end{tabular}\\[0.5em]
\begin{tabular}{cccc}\hline
\rowcolor[gray]{0.8}
5D Brane field
&$\Phi_{\bf 86800}$&$\Phi_{\bf 495}$&$\Phi_{\bf 32}$\\
\hline
$\USp(32)$ &${\bf 86800}$&${\bf 495}$&${\bf 32}$\\
$SO(4,1)$&{\bf 1}&{\bf 1}&{\bf 1}\\
Orbifold BC
&$\left(
\begin{array}{c}
-\\
-\\
\end{array}
\right)$
&$\left(
\begin{array}{c}
+\\
+\\
\end{array}
\right)$
&$\left(
\begin{array}{c}
+\\
+\\
\end{array}
\right)$\\
Spacetime  &$y=0$&$y=0$&$y=0$\\
\hline
\end{tabular}\\[0.5em]
\begin{tabular}{cccc}\hline
\rowcolor[gray]{0.8}
4D Brane field&$\psi_{\overline{\bf 120}}$&
$\psi_{{\bf 16}}^{(b)}$&$\psi_{\overline{\bf 16}}^{(b)}$\\
\hline
$SU(16)$&$\overline{\bf 120}$&${\bf 16}$&$\overline{\bf 16}$\\
$U(1)$  &$0$&$0$&$-1$\\
$SL(2,\mathbb{C})$&$(1/2,0)$&$(1/2,0)$&$(1/2,0)$\\
Spacetime $(y,v)$ &$(0,0)$&$(0,0)$&$(0,0)$\\
\hline
\end{tabular}
\end{center}
\caption{The matter content in
the $\USp(32)$ special GUT on $M^4\times T^2/\mathbb{Z}_2$ is shown.
The representations of $\USp(32)$ and 6D, 5D, 4D Lorentz group, 
the orbifold BCs of 6D bulk fields and 5D brane fields, and 
the spacetime location of 5D and 4D brane fields
are listed.
Orbifold BCs stand for parity assignment \usebox{\BCmatrixA}
 for 6D  fields and \usebox{\BCmatrixB} for 5D fields.
 The orbifold BCs of the 6D $\USp(32)$ gauge field $A_M$ are given in
 Eq.~(\ref{Eq:gauge-field-BCs}).
 $(1/2,0)$ and $(0,1/2)$ of $SL(2,\mathbb{C})$ stand for
 left- and right-handed Weyl fermions, respectively.
 ($a=1,2,3$; $b=1,2,\cdots,12$.) 
\label{tab:USp32-SU16-SO10-matter-content-6D}}
\end{table}

Symmetry breaking consists of three stages in the $\USp(32)$ special GUT
whose $\USp(32)$ is broken into $SU(3)_C\times U(1)_{\rm EM}$:
\begin{enumerate}
 \item Orbifold boundary conditions (BCs) break
       $\USp(32)$ to $SU(16)\times U(1)$.
 \item Nonvanishing vacuum expectation values (VEVs) of three 5D brane
       scalar fields 
       $\Phi_{\bf 86800}$, $\Phi_{\bf 495}$, and $\Phi_{\bf 32}$
       break $SU(16)\times U(1)$ to $G_{\rm SM}$.
 \item A VEV of a zero mode of a 6D bulk
       scalar field $\Phi_{\bf 528}$ breaks
       $G_{\rm SM}$ to $SU(3)_C\times U(1)_{\rm EM}$.
\end{enumerate}
We summarize the above symmetry breaking chain as below:
\begin{align}
 \USp(32)
 \underset{(R)}{\underset{{\rm BCs}}{\longrightarrow}} SU(16)\times U(1)
 \underset{(S)}{\underset{\langle\Phi_{\bf 86800}\rangle\not=0}{\longrightarrow}} SO(10)
 \underset{(R)}{\underset{\langle\Phi_{\bf 32}\rangle,\langle\Phi_{\bf 495}\rangle\not=0}{\longrightarrow}} G_{\rm SM}
 \underset{(R)}{\underset{\langle\Phi_{\bf 528}\rangle\not=0}{\longrightarrow}} SU(3)_C\times U(1)_{\rm EM},
\label{Eq:Symmetry-breaking_USp32-GSM} 
\end{align}
where symmetry breaking chains between $SU(16)$ and
$G_{\rm SM}$ depend on the values of nonvanishing VEVs of
the 5D brane fields
$\Phi_{\bf 86800}$, $\Phi_{\bf 495}$, and $\Phi_{\bf 32}$ and their
associated coupling constants.
The branching rule of $\USp(32)\supset SU(16)\times U(1)$ 
for ${\bf 86800}$ is given by 
\begin{align}
{\bf 86800}=&
({\bf 18240})(0)
\oplus({\bf 14144})(0)
\oplus({\bf 5440})(4)
\oplus({\bf \overline{5440}})(-4)
\oplus({\bf 255})(0)
\oplus({\bf 1})(0)
\nonumber\\
&
\oplus({\bf 21504})(2)
\oplus({\bf \overline{21504}})(-2)
\oplus({\bf 136})(2)
\oplus({\bf \overline{136}})(-2).
\label{branching-rule-86800}
\end{align}
In this letter, we assume the above symmetry breaking and we will not
perform the potential analysis of the scalar fields and determine the
values of their VEVs. 

\subsection{Bosonic sector}

The SM gauge fields are introduced as a part of the 6D $\USp(32)$ bulk
gauge boson $A_{M}$. The 6D $\USp(32)$ bulk gauge boson $A_{M}$ is
decomposed into a 4D gauge field $A_\mu$ and 5th and 6th dimensional
gauge fields $A_y$ and $A_v$, where $A_y$ and $A_v$ are 4D scalar
fields. The orbifold BCs of the 6D $\USp(32)$ gauge field are given by 
\begin{align}
\left(
\begin{array}{c}
A_\mu\\
A_y\\
A_v\\
\end{array}
\right)(x,y_j-y,v_j-v)
&=P_{j{\bf 32}}
\left(
\begin{array}{c}
A_\mu\\
-A_y\\
-A_v\\
\end{array}
\right)(x,y_j+y,v_j+v)
P_{j{\bf 32}}^{-1},\nonumber\\
P_{j{\bf 32}}&=
\left\{
\begin{array}{ll}
I_{32}\ &\mbox{for}\ j=2,3\\
I_{16}\otimes \sigma_3& \mbox{for}\ j=0,1\\
\end{array}
\right.,
\label{Eq:gauge-field-BCs}
\end{align}
where $P_{j{\bf 32}}$ is a projection matrix satisfying 
$(P_{j{\bf 32}})^2=I_{32}$, $I_{n}$ is an $n\times n$ identity
matrix and $\sigma_a(a=1,2,3)$ are the Pauli matrices.
We take a convention for the relation between products $A\otimes B$
and explicit matrix forms given in e.g., Ref.~\cite{Wilczek:1981iz}.
The orbifold BCs $P_2$ and $P_3$ preserve
$\USp(32)$ symmetry, while the orbifold BCs $P_0$ and $P_1$ reduce
$\USp(32)$ to its regular subgroup $SU(16)\times U(1)$.
The orbifold BC that breaks $\USp(32)$ to $SU(16)\times U(1)$ is allowed 
under a $\mathbb{Z}_2$ inner automorphism
\cite{Slansky:1981yr,Hebecker:2001jb}.
Here we take $\USp(32)$ generators as the $32\times 32$ dimensional
Hermitian matrices 
$T_A\otimes I_2$ and $T_{S_a}\otimes\sigma_a$,
where $T_A$ is an anti-symmetric
$16\times 16$ Hermitian matrix and $T_{S_a}$ are symmetric
$16\times 16$ Hermitian matrices \cite{Georgi:1982jb}.
The generators of the
$SU(16)$ subgroup are given by the $32\times 32$ matrices 
$T_A\otimes I_2$ and $T_{S_3}\otimes\sigma_3$ for traceless $T_{S_3}$.
One of the $\USp(32)$ invariant tensors is
$\Omega=I_{16}\otimes i\sigma_2$, which satisfies $U^T\Omega U=\Omega$ 
for ${}^\forall U\in \USp(2n)$.

\begin{table}[t]
\begin{center}
\begin{tabular}{cccc}\hline
 \rowcolor[gray]{0.8}
&\multicolumn{2}{c}{$A_M$}
&\multicolumn{1}{c}{$\Phi_{{\bf 528}}$}
\\
\rowcolor[gray]{0.8}  
$SU(16)\times U(1)$&$M=\mu$&$M=y,v$&\\\hline
$({\bf 255})(0)\oplus({\bf 1})(0)$&
$\left(
\begin{array}{cc}
+&+\\
+&+\\
\end{array}
\right)$&
$\left(
\begin{array}{cc}
-&-\\
-&-\\
\end{array}
\right)$
&$\left(
\begin{array}{cc}
+&+\\
-&-\\
\end{array}
\right)$\\
$({\bf 136})(+2)\oplus({\bf \overline{136}})(-2)$&
$\left(
\begin{array}{cc}
+&+\\
-&-\\
\end{array}
\right)$&
$\left(
\begin{array}{cc}
-&-\\
+&+\\
\end{array}
\right)$
&$\left(
\begin{array}{cc}
+&+\\
+&+\\
\end{array}
\right)$\\
\hline
 \end{tabular}
 \caption{Parity assignments \usebox{\BCmatrixD} of the 4D $SU(16)\times
 U(1)$ gauge and scalar components 
 of the 6D $\USp(32)$ gauge field  $A_M$
 and scalar field $\Phi_{\bf 528}$ are shown.
 $\eta_j(j=0,1,2,3)$ stand for
 $\eta_j=1$ and $\eta_j=-1$ for $A_{M=\mu}$ and $A_{M=y,v}$ regardless
 of $j$; $\eta_0=\eta_1=-1$ and $\eta_2=\eta_3=1$ for $\Phi_{\bf 528}$.
 $T$ represents $T_A\otimes I_2+T_{S_3}\otimes\sigma_3$ and
 $T_{S_1}\otimes\sigma_{1}+T_{S_2}\otimes\sigma_2$ for
 $({\bf 255})(0)\oplus({\bf 1})(0)$ and
 $({\bf 136})(+2)\oplus({\bf \overline{136}})(-2)$, respectively.
 }
\label{Tab:parity-assignment-gauge-528}
\end{center}
\end{table}

We check zero modes of the 6D $\USp(32)$ bulk gauge boson $A_{M}$.
The 4D $\USp(32)$ ${\bf 528}$ gauge field $A_\mu$ have Neumann BCs at
the fixed points $(y_2,v_2)$ and $(y_3,v_3)$, while the 5th and 6th
dimensional gauge fields $A_y$ and $A_v$ have Dirichlet BCs because of
the negative sign in Eq.~(\ref{Eq:gauge-field-BCs}).
On the other hand, since the $\USp(32)$ symmetry is broken to 
$SU(16)\times U(1)$ at the fixed points $(y_0,v_0)$ and $(y_1,v_1)$,
the $SU(16)\times U(1)$ $\left(({\bf 255})(0)\oplus({\bf 1})(0)\right)$
and $\left(({\bf 136})(2)\oplus({\bf \overline{136}})(-2)\right)$ 
components of the 4D gauge field $A_\mu$ 
have Neumann and Dirichlet BCs at the fixed points $(y_0,v_0)$ and
$(y_1,v_1)$, respectively, where 
the branching rule of $\USp(32)\supset SU(16)\times U(1)$ for
${\bf 528}$ is given by Eq.~(\ref{branching-rule-527-USp32-SU16}),
and $(({\bf 255})(0)\oplus({\bf 1})(0))$ and
$(({\bf 136})(2)\oplus({\bf \overline{136}})(-2))$
correspond to the generators
$T_A\otimes I_2+T_{S_3}\otimes\sigma_3$ and 
$T_{S_1}\otimes\sigma_1+T_{S_2}\otimes\sigma_2$, respectively.
The $SU(16)\times U(1)$ 
$\left(({\bf 255})(0)\oplus({\bf 1})(0)\right)$
and $\left(({\bf 136})(2)\oplus({\bf \overline{136}})(-2)\right)$ 
components of the 5th and 6th dimensional gauge 
fields $A_y$ and $A_v$ have Dirichlet and Neumann BCs, respectively.
Thus, since the $SU(16)\times U(1)$ 
$\left(({\bf 255})(0)\oplus({\bf 1})(0)\right)$ components of the 4D
gauge field $A_\mu$ have four Neumann BCs at the four fixed points
$(y_j,v_j) (j=0,1,2,3)$, they have zero modes corresponding to 
4D $SU(16)$ and $U(1)$ gauge fields;
since the other components of $A_\mu$ and any component of $A_y$ and
$A_v$ have four Dirichlet BCs or two Neumann and two Dirichlet BCs at the
four fixed points, they do not have zero modes.
The orbifold BCs reduce $\USp(32)$ to $SU(16)\times U(1)$.
Parity assignments of $A_M$ are shown in
Table~\ref{Tab:parity-assignment-gauge-528}.

The SM Higgs field is introduced as a part of a 6D $\USp(32)$ adjoint
bulk scalar field $\Phi_{\bf 528}$. The 6D $\USp(32)$ bulk scalar
$\Phi_{\bf 528}$ is identified to a 4D scalar field itself.
The orbifold BCs of the 6D $\USp(32)$ scalar field is given by 
\begin{align}
\Phi_{\bf 528}(x,y_j-y,v_j-v)
&=\eta_j P_{j{\bf 32}}
\Phi_{\bf 528}(x,y_j+y,v_j+v)
P_{j{\bf 32}}^{-1},
\label{Eq:scalar-field-BCs}
\end{align}
where $P_{j{\bf 32}}$ is given in Eq.~(\ref{Eq:gauge-field-BCs}) and
$\eta_{0}=\eta_{1}=-1, \eta_{2}=\eta_{3}=1$.

We check zero modes of the 6D $\USp(32)$ adjoint bulk scalar field
$\Phi_{\bf 528}$.
The $\USp(32)$ ${\bf 528}$ scalar field $\Phi_{\bf 528}$ has Neumann
BCs at the fixed points $(y_2,v_2)$ and $(y_3,v_3)$.
Since the $\USp(32)$ symmetry is broken to 
$SU(16)\times U(1)$ at the fixed points $(y_0,v_0)$ and $(y_1,v_1)$,
the $SU(16)\times U(1)$ $\left(({\bf 136})(2)\oplus({\bf \overline{136}})(-2)\right)$ 
and $\left(({\bf 255})(0)\oplus({\bf 1})(0)\right)$
components of the scalar field $\Phi_{\bf 528}$ 
have Neumann and Dirichlet BCs at the fixed points $(y_0,v_0)$ and
$(y_1,v_1)$, respectively.
Therefore, only the $SU(16)\times U(1)$
$\left(({\bf 136})(2)\oplus({\bf \overline{136}})(-2)\right)$ components
have zero modes.
Parity assignments of $\Phi_{\bf 528}$ are shown in
Table~\ref{Tab:parity-assignment-gauge-528}.

The original GUT gauge group $\USp(32)$ must be broken into the SM gauge
symmetry $G_{\rm SM}$ at low energies. It can be realized via 
spontaneous symmetry breaking by introducing 5D $\USp(32)$
${\bf 86800}$, ${\bf 495}$ and ${\bf 32}$ brane scalar
fields $\Phi_{\bf 86800}$, $\Phi_{\bf 495}$, and $\Phi_{\bf 32}$ on the
UV brane ($y=0$), whose orbifold BCs are given by 
\begin{align}
\Phi_{\bf x}(x,v_\ell-v)=&
\eta_{\ell{\bf x}}P_{\ell{\bf x}}\Phi_{\bf x}(x,v_\ell+v),
\end{align}
where $\ell=0,2$,
${\bf x}$ stands for ${\bf 86800}$, ${\bf 495}$ and ${\bf 32}$,
$\eta_{\ell{\bf x}}$ is a positive or negative sign,
and $P_{\ell{\bf x}}$ is a projection matrix.
For $\Phi_{\bf 86800}$, the $SU(16)\times U(1)$ 
$(({\bf 18240})(0)
\oplus({\bf 14144})(0)
\oplus({\bf 5440})(4)
\oplus({\bf \overline{5440}})(-4)
\oplus({\bf 255})(0)
\oplus({\bf 1})(0))$ components
have zero modes,
where ${\bf 5440}$ of $SU(16)$contains ${\bf 1}$ of $SO(10)$;
for $\Phi_{\bf 32}$,
the $SU(16)\times U(1)$ $({\bf 16})(1)$
components have zero modes,
where  ${\bf 16}$ and ${\bf \overline{16}}$ of $SO(10)(\subset SU(16))$
contains ${\bf 1}$ of $SU(5)$;
and for $\Phi_{\bf 495}$, the $SU(16)\times U(1)$ 
$\left(({\bf 255})(0)\right)$ components 
have zero modes,
where ${\bf 255}$ of $SU(16)$ contains $({\bf 1,1})(0)$ of
$G_{\rm SM}(\subset SU(5)\subset SO(10)\subset SU(16))$.
The corresponding parity assignments are realize by choosing
appropriate sings of $\eta_{\ell{\bf x}}$.
The scalar field $\Phi_{\bf 86800}$
is responsible for breaking $(\USp(32)\supset)SU(16)\times U(1)$ to
$SO(10)$; the nonvanishing VEV of the scalar field 
$\Phi_{\bf 32}$ breaks $(\USp(32)\supset)SO(10)$ to $SU(5)$;
the nonvanishing VEV of $\Phi_{\bf 495}$ breaks $(\USp(32)\supset)SU(5)$
to $G_{\rm SM}$.
The above symmetry breaking pattern from $SO(10)$ to $G_{\rm SM}$ is
$SO(10)\to SU(5)\to G_{\rm SM}$.
There is another symmetry breaking pattern 
$SO(10)\to G_{\rm PS}\to G_{\rm SM}$.
In this case,
for $\Phi_{\bf 495}$, the $SU(16)\times U(1)$ 
$\left(({\bf 255})(0)\right)$ components 
have zero modes,
where ${\bf 255}$ of $SU(16)$ contains $({\bf 1,1,1})$ of
$G_{\rm PS}=SU(2)_L\times SU(2)_R\times SU(4)_C(\subset SO(10))$;
for $\Phi_{\bf 32}$,
the $SU(16)\times U(1)$ $({\bf 16})(1)$
components have zero modes,
where  ${\bf 16}$ and ${\bf \overline{16}}$ of
contains $({\bf 1,1})(0)$ of $G_{\rm SM}(\subset G_{\rm PS})$. 
Thus, the nonvanishing VEV of $\Phi_{\bf 495}$ breaks
$(\USp(32)\supset)SO(10)$ to $G_{\rm PS}$.
the nonvanishing VEV of the scalar field 
$\Phi_{\bf 32}$ breaks $(\USp(32)\supset)G_{\rm PS}$ to $G_{\rm SM}$.
The symmetry breaking patterns of $\USp(32)$ broken to $G_{\rm SM}$ are
almost the same as that of $SO(32)$ broken to $G_{\rm SM}$ discussed in
Ref.~\cite{Yamatsu:2017ssg}. 

Symmetry breaking via the non-vanishing VEVs of 
the 5D brane scalar fields 
$\Phi_{\bf 86800}$, $\Phi_{\bf 495}$, and $\Phi_{\bf 32}$ on the UV
brane affects BCs of some components of 6D bulk fields
through 4D brane localized interaction terms
\cite{Hosotani:2017ghg,Hosotani:2017edv}.
For the 6D bulk gauge field $A_M$, the $\USp(32)/G_{\rm SM}$ components
of the 4D components $A_\mu$ have Neumann BCs on the UV brane without the
symmetry breaking effects through the UV brane.
When we take into account the symmetry breaking effects,
the BCs $(y=0)$ become effectively Dirichlet BCs and
originally zero modes acquire masses of $O(m_{\rm KK_5})$, which
depends on coupling constants of corresponding 4D brane
interactions and the values of the VEVs $\langle\Phi_{\bf x}\rangle$s.
In addition, for the 6D bulk scalar field $\Phi_{\bf 528}$,
a 4D brane localized mass term
$\mu\Phi_{\bf 528}\Phi_{\bf 528}$, where
$\mu$ is a mass parameter, is allowed. So,
zero modes of $\Phi_{\bf 528}$ have masses, which are expected to
$O(m_{\rm KK_5})$.
That is, 
all the $SU(16)\times U(1)$
$\left(({\bf 136})(2)\oplus({\bf \overline{136}})(-2)\right)$ component
fields have a common mass.
4D brane localized interaction terms between
the 6D bulk scalar field $\Phi_{\bf 528}$ and
5D brane scalar fields $\Phi_{\bf 86800}$ and $\Phi_{\bf 495}$,
$\kappa\Phi_{\bf 86800}\Phi_{\bf 528}\Phi_{\bf 528}$ and
$\kappa'\Phi_{\bf 495}\Phi_{\bf 528}\Phi_{\bf 528}$, where
$\kappa$ and $\kappa'$ are parameters,
are allowed because a symmetric tensor product of
$\USp(32)$ is given by 
$({\bf 528}\otimes{\bf 528})_{\rm S}=
({\bf 52360})\oplus({\bf 86800})\oplus({\bf 495})\oplus({\bf 1})$.
A term
$\kappa\langle\Phi_{\bf 86800}\rangle\Phi_{\bf 528}\Phi_{\bf 528}$
leads to a mass term
of zero modes of the $SU(16)\times U(1)$
$\left(({\bf 136})(2)\oplus({\bf \overline{136}})(-2)\right)$ component.
The mass of the $(SU(16)\times U(1)\supset) SO(10)\times U(1)$
$\left(({\bf 126})(2)\oplus({\bf \overline{126}})(-2)\right)$ component
is different from
that of the $\left(({\bf 10})(2)\oplus({\bf 10})(-2)\right)$ component
because those Clebsch-Gordan coefficients (CGCs) are different.
Further, a term
$\kappa'\langle\Phi_{\bf 495}\rangle\Phi_{\bf 528}\Phi_{\bf 528}$
leads to a mass term
of zero modes of the $SU(16)\times U(1)$
$\left(({\bf 136})(2)\oplus({\bf \overline{136}})(-2)\right)$ component.
For the $(SU(16)\times U(1)\supset) SO(10)\times U(1)$
$\left(({\bf 10})(2)\oplus({\bf 10})(-2)\right)$ component,
the mass of the $(SU(16)\supset SO(10)\supset)G_{\rm PS}$
$({\bf 1,1,6})$ component
is different from
that of the $(SU(16)\supset SO(10)\supset)G_{\rm PS}$
$({\bf 2,2,1})$ component
because those CGCs are different.
Therefore, we can realize an almost massless $G_{\rm PS}$
$({\bf 2,2,1})$ scalar field, which can be identified as
the SM Higgs field
when we take into account the VEV of $\Phi_{\bf 32}$ that 
breaks $G_{\rm PS}$ to $G_{\rm SM}$, by choosing parameters carefully.

\subsection{Fermionic sector}

\begin{table}[t]
\begin{center}
\begin{tabular}{ccccc}\hline
 \rowcolor[gray]{0.8}
&\multicolumn{2}{c}{$\Psi_{{\bf 32}+}^{(a)}$}
&\multicolumn{2}{c}{$\Psi_{{\bf 32}-}^{(a)}$}
\\
\rowcolor[gray]{0.8}  
$SU(16)\times U(1)$&Left&Right&Left&Right\\\hline
$({\bf 16})(+1)$&
$\left(
\begin{array}{cc}
+&+\\
+&+\\
\end{array}
\right)$&
$\left(
\begin{array}{cc}
-&-\\
-&-\\
\end{array}
\right)$
&$\left(
\begin{array}{cc}
+&-\\
+&-\\
\end{array}
\right)$&
$\left(
\begin{array}{cc}
-&+\\
-&+\\
\end{array}
\right)$\\
$({\bf \overline{16}})(-1)$&
$\left(
\begin{array}{cc}
+&+\\
-&-\\
\end{array}
\right)$&
$\left(
\begin{array}{cc}
-&-\\
+&+\\
\end{array}
\right)$
&$\left(
\begin{array}{cc}
+&-\\
-&+\\
\end{array}
\right)$&
$\left(
\begin{array}{cc}
-&+\\
+&-\\
\end{array}
\right)$\\
\hline
 \end{tabular}
 \caption{Parity assignments \usebox{\BCmatrixC}
 of the 4D $SU(16)\times U(1)$ left- and right-handed Weyl
 fermion components 
 of the 6D $\USp(32)$ ${\bf 32}$ Weyl fermions
 $\Psi_{{\bf 32}\pm}^{(a)}$ $(a=1,2,3)$ are shown.
 }
\label{Tab:parity-assignment-fermion-32}
\end{center}
\end{table}

The SM Weyl fermions are identified with zero modes of 6D $\USp(32)$ 
${\bf 32}$ Weyl bulk fermions, whose orbifold BCs are given by
\begin{align}
\Psi_{{\bf 32}\pm}(x,y_j-y,v_j-v)&=
\eta_{j\pm}\overline{\gamma}
P_{j{\bf 32}}
\Psi_{{\bf 32}\pm}
(x,y_j+y,v_j+v),
\label{Eq:BC-USp(32)-fermion-32}
\end{align}
where the subscript of $\Psi$ $\pm$ stands for 6D chirality,
$\eta_{j\pm}$ is a positive or negative sign satisfying 
$\prod_{j=0}^{3}\eta_{j\pm}=1$,
6D gamma matrices $\gamma^a$ $(a=1,2,\cdots,7)$ 
satisfy $\{\gamma^a,\gamma^b\}=2\eta^{ab}$
($\eta^{ab}=\mbox{diag}(-I_1,I_5)$),
$\overline{\gamma}
:=-i\gamma^5\gamma^6
=\gamma_{\rm 6D}^7\gamma_{\rm 4D}^5$,
$\gamma_{\rm 4D}^5=I_2\otimes \sigma^3\otimes I_2$,
$\gamma_{\rm 6D}^7=I_4 \otimes \sigma^3$.
In our notation, a 6D Dirac fermion $\Psi_{\rm D}^{\rm 6D}$
and 6D positive and negative Weyl fermions
$\Psi_{\pm}^{\rm 6D}:=P_{\pm}^{\rm 6D}\Psi_{\rm D}^{\rm 6D}$
($P_{\pm}^{\rm 6D}:=(1\pm\gamma_{\rm 6D}^7)/2$)
can be expressed by using 4D left- and right-handed Weyl fermions
$\psi_{L/R\pm}^{\rm 4D}(=P_{L/R}^{\rm 4D}\psi_{\rm D}^{\rm 4D})$
($P_{L/R}^{\rm 4D}:=(1\pm\gamma_{\rm 4D}^5)/2$),
where subscripts $L/R$ and $\pm$ stand for
4D and 6D chiralities, respectively:
\begin{align}
 &\Psi_{\rm D}^{\rm 6D}
 :=
 \left(
 \begin{array}{c}
  \psi_{R+}^{\rm 4D}\\
  \psi_{L+}^{\rm 4D}\\
  \psi_{R-}^{\rm 4D}\\
  \psi_{L-}^{\rm 4D}\\
 \end{array}
 \right),\ \
 \Psi_{+}^{\rm 6D}
 =P_{+}^{\rm 6D}\Psi_{\rm D}^{\rm 6D}
 =\left(
 \begin{array}{c}
  \psi_{R+}^{\rm 4D}\\
  \psi_{L+}^{\rm 4D}\\
  0\\
  0\\
 \end{array}
 \right),\ \
 \Psi_{-}^{\rm 6D}
 =P_{-}^{\rm 6D}\Psi_{\rm D}^{\rm 6D}
 = \left(
 \begin{array}{c}
  0\\
  0\\
  \psi_{R-}^{\rm 4D}\\
  \psi_{L-}^{\rm 4D}\\
 \end{array}
 \right).
\end{align}
We omit the superscripts such as 6D and 4D below.

Here we check zero modes of a 6D $\USp(32)$ ${\bf 32}$ positive Weyl
fermion with orbifold BCs $\eta_{j}=-1$. 
In this case, only a 4D $SU(16)\times U(1)$ $({\bf 16})(1)$ 
left-handed Weyl fermion has a zero mode.
This is because only the left-handed Weyl fermion
component has all the Neumann BCs at all the fixed points $(y_j,v_j)$.

To realize three generations of the SM chiral fermions, we need to
introduce three 6D $\USp(32)$ ${\bf 32}$ positive Weyl fermions. 
However, only 6D $\USp(32)$ ${\bf 32}$ positive Weyl fermions suffer
from 6D gauge anomalies. It is known that the 6D gauge anomalies can be
canceled out by introducing additional 6D $\USp(32)$ ${\bf 32}$ negative
Weyl fermions with different BCs as discussed in
Refs.~\cite{Yamatsu:2017sgu,Yamatsu:2017ssg,Yamatsu:2018fsg}.
Parity assignments of $\Psi_{\bf 32}^{(a)}$ are shown in
Table~\ref{Tab:parity-assignment-fermion-32}

We check contributions to 4D brane anomalies from the above 6D Weyl
fermion sets.
At two fixed points $(y_j,v_j) (j=2,3)$, there is no 4D pure $\USp(32)$
gauge anomaly because any 4D anomaly coefficient of $\USp(32)$ is zero.
At the other two fixed points $(y_j,v_j) (j=0,1)$, there can be
4D pure $SU(16)$, pure $U(1)$, mixed
$SU(16)-SU(16)-U(1)$ and mixed $\mbox{grav.}-\mbox{grav.}-U(1)$
anomalies.
At a fixed point $(y_1,v_1)$, the anomalies generated from the 6D
$\USp(32)$ ${\bf 32}$ positive and negative Weyl fermions are canceled
each other.
Finally,  at the other fixed point $(y_0,v_0)$, the 6D $\USp(32)$ 
${\bf 32}$ positive and negative Weyl fermions generate 
4D pure $SU(16)$, pure $U(1)$, mixed
$SU(16)-SU(16)-U(1)$ and mixed $\mbox{grav.}-\mbox{grav.}-U(1)$
anomalies.
As the same discussion in Ref.~\cite{Yamatsu:2017ssg},
the 4D gauge anomalies can be canceled out by introducing 
4D brane Weyl fermions in appropriate representations of
$SU(16)\times U(1)$ shown in
Table~\ref{tab:USp32-SU16-SO10-matter-content-6D}.

\section{Summary and discussion}
\label{Sec:Summary-discussion}

In this letter, we proposed a $\USp(32)$ special GUT by using a
special breaking $SU(16)$ to $SO(10)$. Zero modes
of a 6D $\USp(32)$ ${\bf 32}$ Weyl fermion are identified with
one generation of quarks and leptons;
the 6D $\USp(32)$ and the 4D $SU(16)\times U(1)$ gauge anomalies on the
fixed points allow a three generation model of quarks and leptons in a
6D framework; as in the $SU(16)$ special GUT \cite{Yamatsu:2017sgu}, 
exotic chiral fermions do not exist.
The SM Higgs scalar field is introduced as a part of a 6D $\USp(32)$
adjoint bulk scalar field. Unfortunately, to realize an almost massless
SM Higgs scalar field and the other massive scalar fields, the
fine-tuning is inevitable.

In the $\USp(32)$ special GUT, the SM fermions and the SM Higgs scalar
are embedded into zero modes of the 6D bulk fields, so 
the masses and mixing matrices are given by the overlap integral of
the wave functions of zero modes of the 6D bulk fermion fields
$\Psi_{{\bf 32}+}^{(a)}$ and 
the 6D bulk scalar field  $\Phi_{{\bf 528}}$.
For an $SO(11)$ gauge-Higgs GUT model in the 6D hybrid warped space
\cite{Hosotani:2017ghg,Hosotani:2017edv},
hierarchical masses of fermions are realized by taking the values of
bulk vector masses in the 6D hybrid warped space, which corresponds to
bulk scalar masses in the 5D RS space.
To explain tiny neutrino masses by a see-saw mechanism
\cite{Minkowski:1977sc,Yanagida:1979as,GellMann:1980vs,Mohapatra:1979ia,Schechter:1980gr,Mohapatra:1986aw,Mohapatra:1986bd},
additional $SO(11)$ singlet brane
fermions are introduced on the UV brane, which satisfy a symplectic
Majorana condition \cite{Kugo:1982bn,Mirabelli:1997aj}.
The additional brane fermions, bulk fermions, and the $SO(11)$ breaking
brane scalar fields lead to additional brane mass terms on the UV
brane. The mass terms generate tiny neutrino masses. A similar
discussion is given in Ref.~\cite{Hasegawa:2018jze}.
Also, a mixing matrix in the quark sector, the Cabibbo-Kobayashi-Maskawa
(CKM) mixing matrix \cite{Cabibbo:1963yz,Kobayashi:1973fv},
is introduced by brane interaction terms
\cite{Cacciapaglia:2007fw,Adachi:2010cc,Adachi:2011tn,Adachi:2011cb,Funatsu:2019fry}.
We will leave further discussions for future studies.

We comment on the possibility of gauge-Higgs unification in a
$\USp(32)$ special GUT.
As is discussed in the letter, a 6D bulk scalar field in the adjoint
representation of $\USp(32)$ contains the SM Higgs scalar component, so
the 4D scalar component of a 6D $\USp(32)$ bulk gauge field contains
also the SM Higgs scalar component, but
the BCs in Eq.~(\ref{Eq:gauge-field-BCs}) do not allow zero modes of the
4D scalar component.
Instead of the BCs in Eq.~(\ref{Eq:gauge-field-BCs}), if we take the
BCs $P_{j{\bf 32}}=I_{16}\otimes\sigma_3$ $(j=0,1,2,3)$,
the 4D scalar fields of the $SU(16)\times U(1)$
$\left(({\bf 136})(2)\oplus({\bf \overline{136}})(-2)\right)$ component
have zero modes, where a part of the 4D scalar fields can be identified
the SM Higgs scalar field. However, the BCs do not allow three
chiral generations of the SM fermions because
each 6D bulk Weyl fermion in a $\USp(32)$ defining
representation {\bf 32} has two or no zero modes of the 4D
left-handed or right-handed Weyl fermions in a $SO(10)$ spinor
representation ${\bf 16}$ depending on each parity assignment.
We need some additional ideas to construct unified models satisfying
both $\USp(32)$ grand unification and gauge-Higgs unification.
We will leave further discussions for future studies.

Finally, we comment on symmetry breaking. 
Many people vaguely believe that symmetry groups are broken to only
regular subgroups, not to special subgroups. 
However,
the symmetry breaking of $SU(n)$ to its special subgroups such as
$SO(n)$ and $\USp(2[n/2])$ are known to be realized by
a nonvanishing VEV of a fundamental
scalar field in rank-2 symmetric and anti-symmetric tensor
representations of $SU(n)$ 
\cite{Li:1973mq,Meljanac:1982rc,Abud:1984xn},
a nonvanishing VEV of a composite scalar field made by fermion
pair condensation in fundamental and rank-2 anti-symmetric tensor
representations of $SU(n)$ \cite{Kugo:2019isl},
orbifold boundary conditions (BCs) by $\mathbb{Z}_2$ outer
automorphisms on $S^1/\mathbb{Z}_2$ orbifold space \cite{Hebecker:2001jb}.
Also, other symmetry groups such as $SO(n)$ and $E_6$ broken to their
special subgroups are discussed
in Ref.~\cite{Li:1973mq} for fundamental scalar fields;
in Refs.~\cite{Kugo:1994qr,Kugo:2019wge} for composite scalar fields;
in Ref.~\cite{Hebecker:2001jb} for $\mathbb{Z}_2$ orbifold space.

\section*{Acknowledgments}

The author would like to thank Kentaro Kojima and Kenji Nishiwaki
for many valuable comments. 
This work was supported in part by the MEXT/JSPS KAKENHI Grant Number
JP18H05543 and JP19K23440.

{%\small
\bibliographystyle{utphys} 
\bibliography{../../arxiv/reference}
}
\end{document}